\documentclass[english]{IEEEtran}
\usepackage[T1]{fontenc}
\usepackage[latin9]{inputenc}
\usepackage{array}
\usepackage{amstext}
\usepackage{graphicx}

\makeatletter

\providecommand{\tabularnewline}{\\}

\usepackage{graphicx}
\usepackage{dcolumn}
\usepackage{bm}
\usepackage{cite}
\usepackage{amsmath,amssymb,amsfonts}
\usepackage{algorithmic}
\usepackage{textcomp}
\def\BibTeX{{\rm B\kern-.05em{\sc i\kern-.025em b}\kern-.08em
    T\kern-.1667em\lower.7ex\hbox{E}\kern-.125emX}}

\makeatother

\usepackage{babel}
\begin{document}

\title{Strongly correlated proton-doped perovskite nickelate memory devices}

\author{Koushik Ramadoss$^{1}$, Fan Zuo$^{1}$, Yifei Sun$^{1}$, Zhen Zhang$^{1}$,
Jianqiang Lin$^{2,3}$, Umesh Bhaskar$^{4}$, SangHoon Shin$^{4}$,
Muhammad Ashraful Alam$^{4}$, Supratik Guha$^{2,3}$, Dana Weinstein$^{4}$
and Shriram Ramanathan$^{1}$\thanks{The authors acknowledge financial support from ARO W911NF-16-1-0289
and AFOSR FA9550-16-1-0159. This work was done in part at the Birck
Nanotechnology Center, Purdue University.} \thanks{K Ramadoss, F Zuo, Y Sun, Z Zhang and S Ramanathan are with the School
of Materials Engineering, Purdue University, West Lafayette, IN 47907
USA (e-mail: koushikr.in@gmail.com).} \thanks{U Bhaskar, D Weinstein, S Shin and M Alam are with the School of Electrical
and Computer Engineering, Purdue University, West Lafayette, IN 47907
USA.}\thanks{J Lin and S Guha are with the Center for Nanoscale Materials, Argonne
National Laboratory, Lemont, IL 60439 USA and Institute for Molecular
Engineering, University of Chicago, Chicago, IL 60615 USA}}
\maketitle
\begin{abstract}
We demonstrate memory devices based on proton doping and re-distribution
in perovskite nickelates (RNiO$_{3}$, $\mathrm{R=Sm,}\mathrm{Nd}$)
that undergo filling-controlled Mott transition. Switching speeds
as high as $30$~ns in two-terminal devices patterned by electron-beam
lithography is observed. The state switching speed reported here are
$\sim300$$\times$ greater than what has been noted with proton-driven
resistance switching to date. The ionic-electronic correlated oxide
memory devices also exhibit multi-state non-volatile switching. The
results are of relevance to use of quantum materials in emerging memory
and neuromorphic computing.
\end{abstract}


\section{Introduction}

Resistive switching in oxides has the potential to enable high speed,
densely scaled non-volatile memories for storage applications and
emerging neuromorphic computing circuits~\cite{ielmini2015resistive}.
These devices that form a part of RRAM technology are also being considered
to replace NAND flash in the near future. RRAM usually involves a
MIM structure whose resistance can be switched between low resistance
state (LRS) and high resistance state (HRS) by application of a bias
voltage. Many of these devices function based on the principle of
local conductive filament formation or interface switching involving
movement of oxygen vacancies in defective wide gap insulating oxides~\cite{akinaga2010resistive,ielmini2015resistive}.
At the same time, there is a need to innovate in new materials and
physical phenomena for state switching, control of volatility and
electric-field driven operation of memory devices without temperature
constraints. In this work, we use proton-doped SmNiO$_{3}$ (SNO)
and NdNiO$_{3}$ (NNO) as model strongly correlated quantum material
systems to demonstrate high speed, non-volatile memory that do not
require any forming voltage for its operation. Perovskite nickelates
display collective properties like thermally-driven metal-insulator
transition and anti-ferromagnetism and are being explored for various
technologies such as sensing, photovoltaics and resistance switching~\cite{middey2016physics,catalano2018rare}.
These materials are weakly insulating (gap of the order of $\sim100$~meV
in the ground state). A large resistance change occurs due to proton-electron
coupled doping from filling-control of Ni $\textrm{e}_{g}$ orbitals
in the nickelate leading to Mott transition that is distinct from
any thermal phase transition~\cite{shi2014colossal}. This doping
process is based on splitting of hydrogen into a proton and electron
at a catalytic electrode like Pd, Pt resulting in the modification
of Ni orbital occupancy that leads to an opening of large band gap
($3$~eV) when the Ni site occupancy changes from $\textrm{e}_{g}^{1}$
to $\textrm{e}_{g}^{2}$. The protons reside in interstitial sites
in the perovskite lattice. By applying voltage pulses, it is possible
to realize distinct resistance states by spatially varying the dopant
concentration with application of bias and the process is reversible.
Orbital occupancy control via external electric field serves as the
operational mechanism for these devices. While a mainstay topic of
interest in physics, much work remains in exploring the potential
of Mott transition in electronic devices. In this work we present
the first demonstration of high speed non-volatile memory devices
using nickelates, notably $300\times$ improvement in switching speed
of protonic switches over what is reported in literature. 

\section{Device fabrication}

\begin{figure}[th]
\begin{centering}
\includegraphics[width=0.39\textwidth]{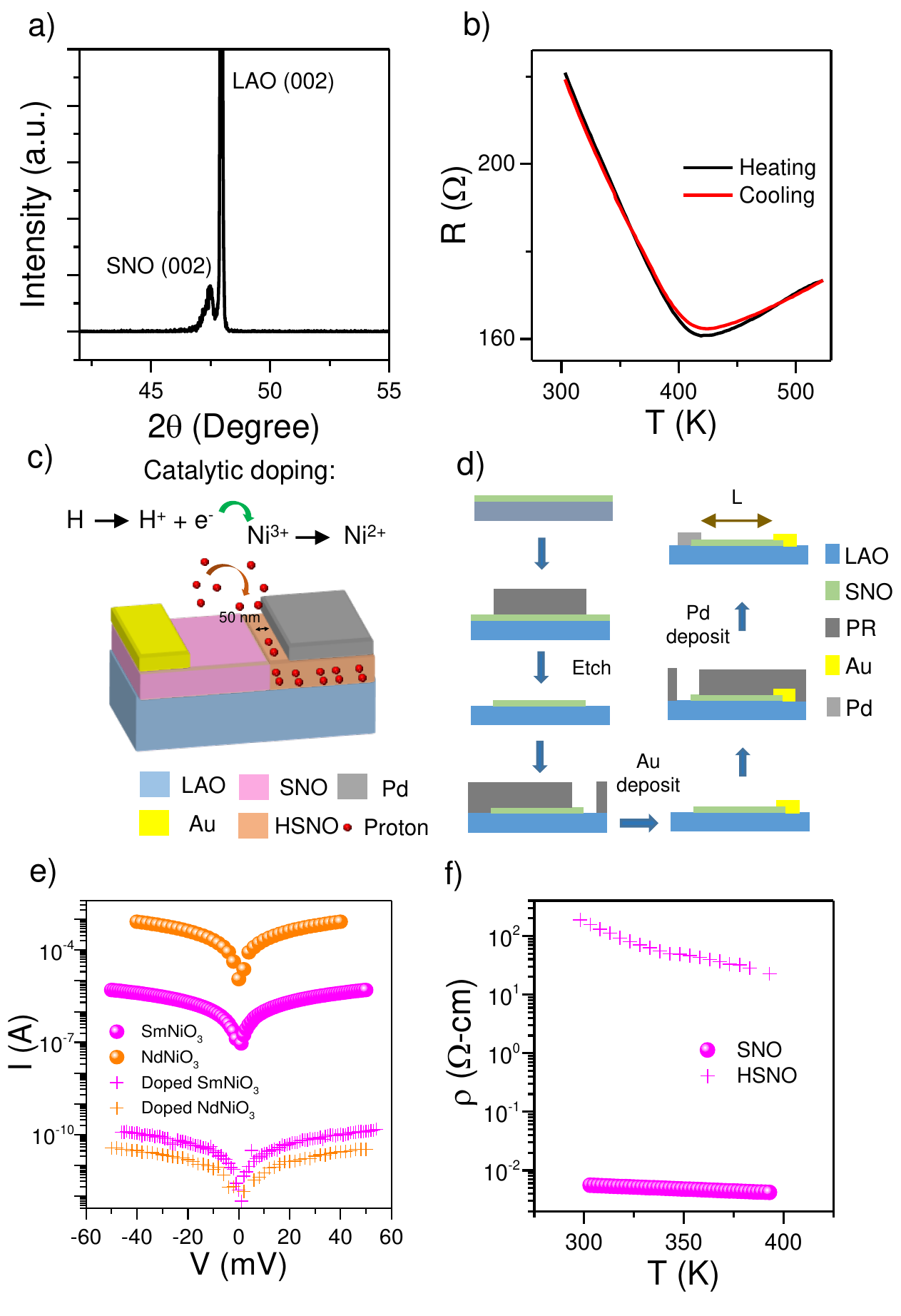}
\par\end{centering}
\caption{a) XRD of a thin film of SNO ($50$~nm) grown on LAO substrate. b)
Thermal insulator-metal transition in thin film of un-doped SNO/LAO
showing the expected transition around $402$~K. c) Mott memory device
with electrode-selective proton incorporation by catalytic doping
at Pd/SNO interface. d) Process flow diagram for fabrication of memory
device with asymmetric electrode configuration (PR : Photoresist).
For symmetric devices, both electrodes are chosen to be Pd. e) Typical
IV curves for pristine (SNO \& NNO) and protonated regions (Doped
SNO \& Doped NNO) showing a large increase in the resistance due to
proton incorporation . f) Temperature dependence of resistance of
a typical device (encapsulated with $15$~nm SiO$_{2}$ passivation
layer) before and after doping (HSNO) indicating stability of protons
in the lattice. }
 \label{fig:f1}
\end{figure}
Thin films of SNO ($50$~nm) were grown on LaAlO$_{3}$ (LAO) substrates
by RF co-sputtering from Sm and Ni targets and subsequently annealed
under high pressure oxygen ($1500$~PSI) at $500{^\circ}$C to form
the crystalline phase. The X-Ray diffraction of a pristine SNO film
is shown in Fig. \ref{fig:f1}a. These un-doped films show a thermal
insulator-metal transition around $402$~K as shown in Fig. \ref{fig:f1}b.
The thin films have a perovskite structure with optimal nickel valency
of Ni$^{3+}$ and can be catalytically doped with protons using metal
electrodes (like Pt, Pd) on top of SNO and annealing in forming gas
(Fig. \ref{fig:f1}c). During this process, the hydrogen breaks into
proton and electron at the metal/SNO interface with the electron anchoring
in the nickel orbitals changing the valence state from Ni$^{3+}$
to Ni$^{2+}$, thus producing a strongly correlated electronic state.
The process flow for fabricating these devices is shown in Fig. \ref{fig:f1}d.
The fabrication of these memory cells starts with patterning pristine
SNO films into rectangular bars using photolithography and then devices
with either asymmetric electrodes (Ti/Au ($10/100$~nm) and Pd ($100$~nm))
or symmetric electrodes (Pd ($100$~nm)) were patterned to define
a channel of width $100$~$\mu$m and length $5$~$\mu$m. To fabricate
sub-micron devices with channel length of few $100$~nm, electron
beam lithography was used for patterning. Protons are then doped into
the device by annealing them under forming gas at $100{^\circ}$C
for about $15$~minutes in a home built furnace. The dopants are
localized close to the catalytic Pd electrode and can re-distribute
under electric bias thereby changing the resistance of the channel.
The moderate temperature annealing conditions ensures the dopants
are incorporated into the lattice in a non-volatile manner.

To understand the electrical behavior before and after doping, IV
characteristics of the devices (both SmNiO$_{3}$ \& NdNiO$_{3}$)
are shown in Fig. \ref{fig:f1}e. The presence of Ni$^{2+}$ after
proton-electron doping exhibits strong electronic correlations~\cite{shi2014colossal}
leading to a massive increase in resistance of the material by several
orders of magnitude (Fig. \ref{fig:f1}e). This is due to a large
increase in bandgap of the order of $3$~eV due to localization upon
half-filling (doubly occupied $\textrm{\ensuremath{e_{g}}}$ state).
The extent of resistance modulation is governed by the voltage-driven
diffusion process of protons (that reside in interstitial sites in
the perovskite lattice) and so can be varied. This is also the elementary
mechanism for designing multiple resistance levels in a memory cell.
Fig. \ref{fig:f1}f shows the dependence of resistance with temperature
for both the pristine (SNO) and proton doped films (HSNO), passivated
by $15$~nm SiO$_{2}$ layer, indicating the stability of protons
in these materials. The experiment for resistive swtiching is performed
in a shielded probe station equipped with Keithley 4200 Semiconductor
parameter analyzer with a pulse generation unit and a remote amplifier
switch that connects between pulse generator and source measuring
unit (SMU). The resistance state of the devices is read from the IV
response by restricting the voltage levels to within $\pm50$~mV. 

\section{Results and Discussion}

\begin{figure}[tbh]
\begin{centering}
\includegraphics[width=0.45\textwidth]{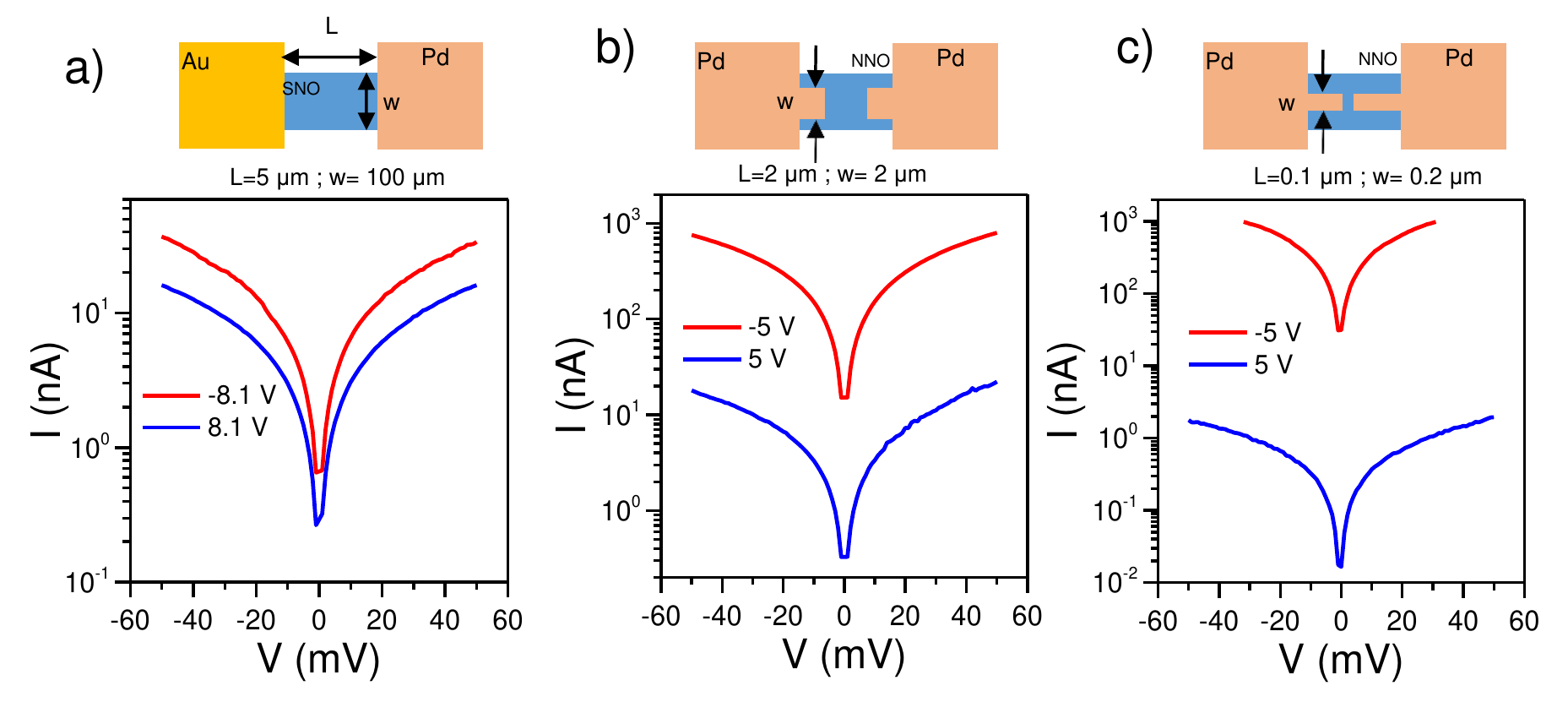}
\par\end{centering}
\caption{a) IV curve of the device ($L=5$~$\mu$m) after application of pulses
(amplitude =$\pm8.1$~V, duration = $130$~ns) along with the device
schematic. b) IV curve of another device ($L=2$~$\mu$m) after application
of pulses (amplitude =$\pm5$~V, duration = $30$~ns) along with
the device schematic. c) IV curve of a sub-micron device ($L=0.1$~$\mu$m)
after application of pulses (amplitude =$\pm5$~V, duration = $100$~ns)
along with the device schematic.}
 \label{fig:f2}
\end{figure}
 The IV curves of a typical cell corresponding to HRS (blue curve)
and LRS (red curve) is shown in Fig.~\ref{fig:f2}a-c for devices
with different channel dimensions along with their schematics. The
device, initially in HRS is switched to LRS by application of a pulse
of negative amplitude. While the micro scale device ($L=5$~$\mu$m)
showed switching from $\textrm{\ensuremath{\textrm{R}_{High}}}=3.3$~M$\Omega$
to $\textrm{\ensuremath{R_{Low}}}=$$1.5$~M$\Omega$ for a pulse
of amplitude $-8.1$~V and duration of about $130$~ns, the sub
micron device ($\mathrm{L}=0.1$~$\mu$m) showed a larger change
in resistance from $\textrm{\ensuremath{\textrm{R}_{High}}}=29.2$~M$\Omega$
to $\textrm{\ensuremath{R_{Low}}}=$$31.7$~k$\Omega$ for a pulse
of amplitude $-5$~V and duration of about $100$~ns. The resistance
switching ratio ($\textrm{\ensuremath{R_{High}}}/\textrm{\ensuremath{R_{Low}}}$)
is about $920$ for a sub-micron device as compared to $2.2$ in a
micro scale device. In a device with intermediate length ($\mathrm{L}=2$~$\mu$m),
the switching ratio is about $43$ (Fig.~\ref{fig:f2}b ). Note that
in micron-scale devices, several volts (over longer duration) are
required to provide sufficient driving force, while this scales with
the junction dimensions as discussed in the next paragraph. We have
also observed resistive switching at $30$~ns timescale in our devices
as shown in Fig.~\ref{fig:f2}b. Additional experiments to characterize
the nature of switching ie. the effects of applied pulse voltage and
duration and test for non-volatility have been performed on representative
devices. 
\begin{figure}[tbh]
\begin{centering}
\includegraphics[width=0.4\textwidth]{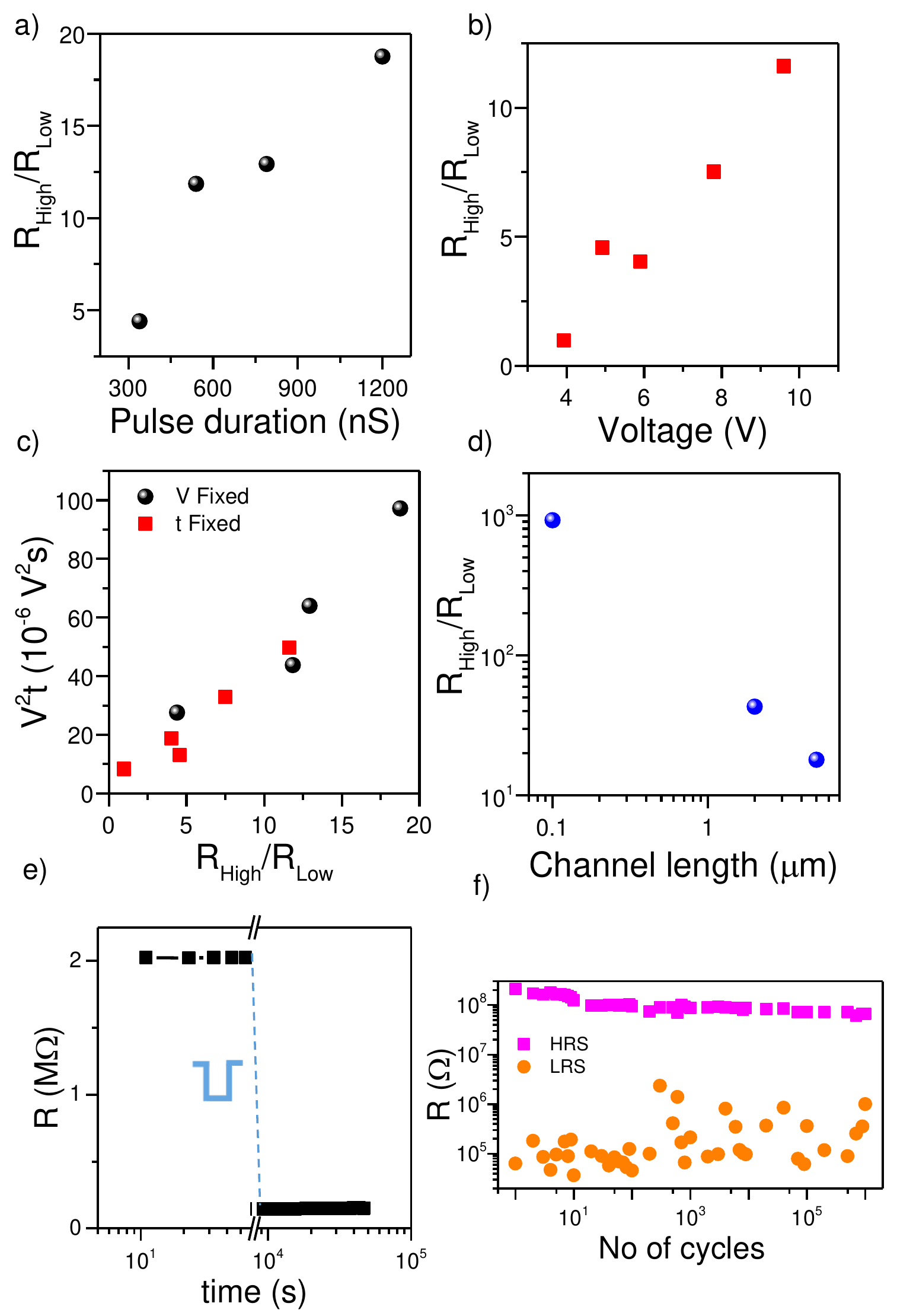}
\par\end{centering}
\caption{a) Switching ratio as a function of different pulse duration at $9.1$~V
in a RRAM cell. b) Switching ratio as a function of applied pulse
voltages with pulse duration of $540$~ns. The switching ratio scales
with pulse width and duration due to larger displacement of protons
in the channel. c) Scaling of V$^{2}$t as a function of switching
ratio indicates that larger switching ratio is acieved by supplying
larger energy to the devices. d) Scaling in Mott memory: Maximum value
of switching ratio for devices as a function of channel length. e)
Non-volatility test of memory device. The LRS is retained for more
than $11$ hours. f) Endurance test of cell (passivated with $15$~nm
SiO$_{2}$) switched using pulses of duration $100$~ns and amplitude
$\pm5$~V. The device is stable for $>10^{6}$ cycles.}
 \label{fig:f3}
\end{figure}
\begin{table*}[tp]
\caption{Comparison of the relevant metrics reported for proton-doped resistive
switching devices}
\label{tab1}
\centering{}%
\begin{tabular}{|>{\centering}m{2cm}|>{\centering}m{3cm}|>{\centering}m{2cm}|>{\centering}m{2cm}|>{\centering}m{2cm}|>{\centering}m{2cm}|>{\centering}m{2cm}|}
\hline 
\noalign{\vskip0.05cm}
\textbf{Material} & \textbf{Doping method} & \textbf{Switching time scale} & \textbf{Channel length} & \textbf{Switching voltage} & \textbf{Switching Ratio} & \textbf{Ref}\tabularnewline[0.05cm]
\hline 
\noalign{\vskip0.05cm}
WO$_{3}$ & Diffusion (hygroscopic film) & $\sim2$ minutes & $>500$~$\mu$m & $10$~V & $10$ & ~\cite{thakoor1990solid}\tabularnewline[0.05cm]
\hline 
\noalign{\vskip0.05cm}
NdNiO$_{3}$ & Catalytic doping & $\sim5$~milli seconds & $\sim700$~$\mu$m & $1$~V & $10$ & ~\cite{oh2016correlated}\tabularnewline[0.05cm]
\hline 
\noalign{\vskip0.05cm}
SiO$_{x}$ & Proton exchange reaction & $\sim10$~micro seconds & $2$~$\mu$m & $\sim5$~V & $100$ & ~\cite{chang2016demonstration}\tabularnewline[0.05cm]
\hline 
NdNiO$_{3}$ & Catalytic doping & $\sim30$~nano seconds & $0.1$~$\mu$m & $5$~V & $920$ & \textbf{This work}\tabularnewline[0.05cm]
\hline 
\end{tabular}
\end{table*}

We have measured the switching ratio as a function of pulse duration
keeping the amplitude constant ( $\pm9.1$~V) (Fig.~\ref{fig:f3}a)
and as a function of pulse voltage keeping the pulse duration constant
($540$~ns) (Fig.~\ref{fig:f3}b). We observe a linear scaling of
switching ratio as the pulse duration and voltage increase. This seems
to suggest that the process of switching is governed by energy supplied
to displace the protons. To confirm this, we plot $\textrm{\textrm{\ensuremath{\textrm{V}^{2}}t}}$
as a function of switching ratio (Fig.~\ref{fig:f3}d), where we
see that the data in plots in Fig.~\ref{fig:f3}a,b collapse on to
each other and the linear scaling shows that the switching ratio for
the device increases with applied energy. The switching ratio also
scales with the device dimensions as shown in Fig.~\ref{fig:f3}d
where the maximum value increases as the channel length is reduced.
We also note that protons possessing smaller ionic radii than oxygen
ions or cations cause less distortions in the underlying lattice.
This may translate eventually into lower activation energy for ion-driven
switching.

\begin{figure}[tbh]
\begin{centering}
\includegraphics[width=0.36\textwidth]{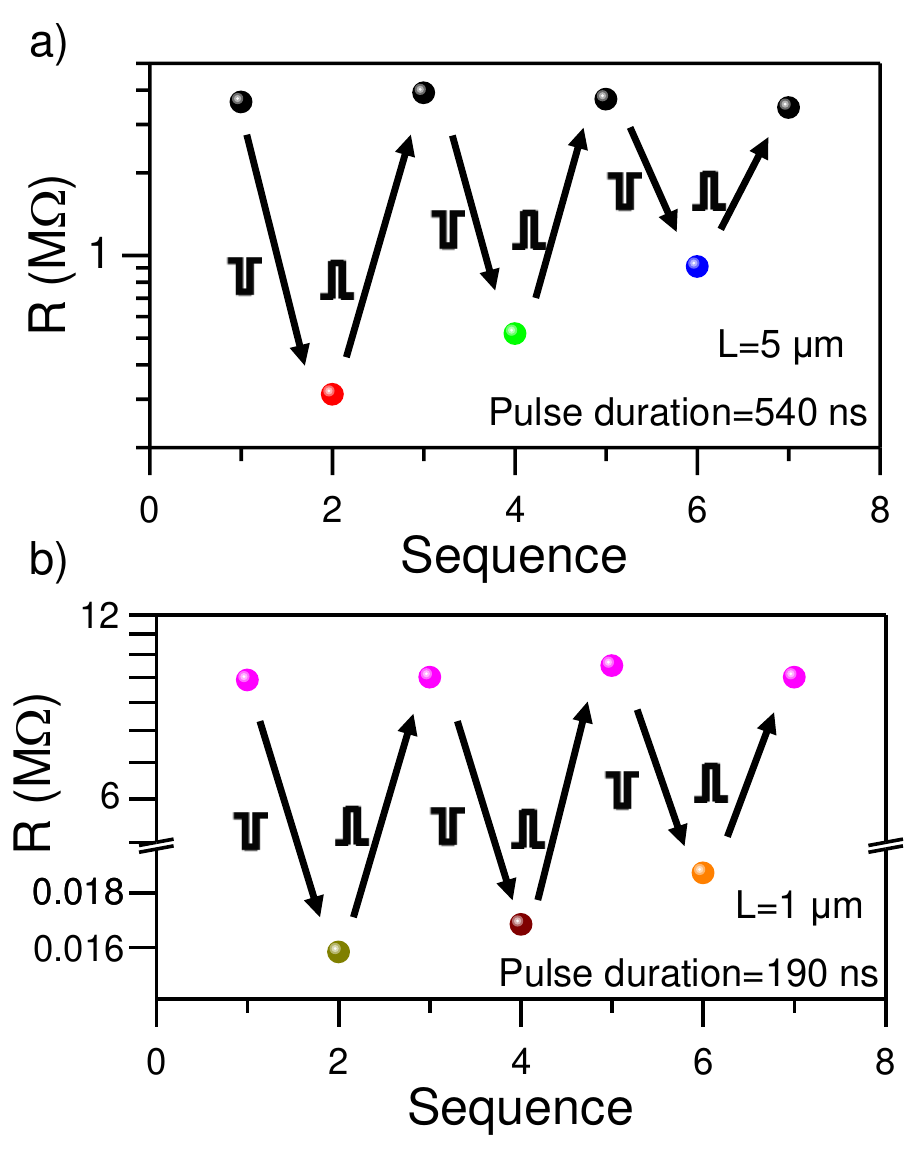}
\par\end{centering}
\caption{a) Multi-state switching demonstrated in a device with $L=5$$\mu$m.
Re-distribution of protons across the device by application of different
voltage bias ($\mathrm{V}=-5.9~\mathrm{V},-7.8~\mathrm{V},-9.6~\mathrm{V}$
for LRS and $\mathrm{V}=9.6$~V for HRS, pulse duration $=540$~ns)
enables multi-states. The resistance ratio for a given switching time
can be increased in smaller gap devices by extending the regime of
electron doping in the channel. b) As an example multi-state switching
is demonstrated in a device with $\mathrm{L}=1$~$\mu$m ( $\mathrm{V}=-8~\mathrm{V},-9~\mathrm{V},-10~\mathrm{V}$
for LRS and $10$~V for HRS, pulse duration $=190$~ns).}
 \label{fig:f4}
\end{figure}
We then evaluate non-volatile behavior in the cells (Fig.~\ref{fig:f3}e).
The device (initially in HRS) is switched to LRS and the resistance
is monitored continuously in the LRS. The change in the resistance
(in LRS) is less than $15$~\% measured over $11$~hours. In Fig.~\ref{fig:f3}f,
we show results of an endurance test performed on a device with a
switching time of $100$~ns and pulse voltage of $\pm5$~V and find
that the device is stable over $10^{6}$ cycles which is an encouraging
initial result. Future studies are necessary to better understand
the reliability and endurance properties of such Mott devices. A comparison
with literature reports on proton doped oxides~\cite{thakoor1990solid,oh2016correlated,chang2016demonstration}
exhibiting resistive switching (Tab.~\ref{tab1}) indicate that our
devices show higher speed and switching ratio, thus suggesting that
filling-controlled Mott transition based nickelate memory is an interesting
candidate among new materials being considered for emerging memory
technologies.

We have further studied multi-level operation by switching to various
resistance states by varying the applied pulse voltage while keeping
the pulse duration fixed at $540$~ns (Fig.~\ref{fig:f4}a). The
device is initially in HRS ($\textrm{R}\sim2.5$~M$\Omega$) and
switched to LRS ($\textrm{R}\sim300$~k$\Omega$) with $-9.6$~V
pulse. The device is then taken back to its HRS by $9.6$~V pulse.
Then the device is sequentially taken through various resistance states
in sequence by tuning the pulse amplitude. Fig.~\ref{fig:f4}b shows
the multi-level switching of another device ($\mathrm{L}=1$~$\mu$m)
with pulse duration of $190$~ns.  

\section{Conclusions }

We have demonstrated pulsed voltage-driven resistance switching in
proton-doped perovskite nickelates. Under voltage bias, the protons
rapidly re-distribute leading to distinct resistance states. The results
suggest the physical phenomena of orbital filling-controlled Mott
transition in correlated oxides can be relevant for emerging memory
devices.

\bibliographystyle{IEEEtran}
\bibliography{E:/Data/Purdue/Paper/ReRAM/NVM}

\end{document}